\documentclass{camera}
\usepackage{epsfig}

\begin{document}

%%%%%%%%%%%%%%%%%%%%%%%%%%%%%%%%%%%%%%%%%%%%%%%%%%%%%%%%%%%%%%%%%%%%
\title{IN SEARCH FOR PHYSICS BEYOND THE STANDARD MODEL AT TEVATRON}
%%%%%%%%%%%%%%%%%%%%%%%%%%%%%%%%%%%%%%%%%%%%%%%%%%%%%%%%%%%%%%%%%%%%
\author{C. Pagliarone$^{1}$, E. Vataga$^{2}$\\
$^{1}$INFN of Pisa\\ $^{2}$INFN of Pisa \& Moscow State
University\\ {\it via Livornese, 1291 - 56010 S. Piero a Grado
(PI) - ITALY}\\ {\it e-mail: pagliarone@fnal.gov,
vataga@fnal.gov}\\ (ON THE BEHALF OF THE CDF\& D\O\
COLLABORATION)}

\maketitle \vspace{-0.9cm}

\begin{abstract}
We present the most recent results of searches for physics beyond
the Standard Model using the CDF and the D$\not$O detector at the
Fermilab Tevatron Collider. All results shown correspond to
analysis performed using the past 1992-1996 Fermilab Tevatron run
I data (roughly $110$ $pb^{-1}$ per each experiment). In
particular we report on searches for scalar top and scalar bottom
particles together with other {\it classic} Supersymmetry
analysis. Results from non Standard Model Higgs searches are also
summarized.
\end{abstract}
% We summarized the results of non-SM Higgs searches.
% Prospects of the Tevatron experiments in Run II are also discussed.}
%%%%%%%%%%%%%%%%%%%%%%%%%%%%%%%%%%%%%%%%%%%%%%%%%%%%%%%%%%%%%%%%%%%%%%%%%%%%%%%%%%%%

%%%%%%%%%%%%%%%%%%%%%%%%%%%%%%%%%%%%%%%%%%%%%%%%%%%%%%%%%%%%%%%%%%%%%%%%%%%%%%%%%%%%%
\section{Introduction}
Although, at present, the Standard Model (SM) provides a remarkably successful description of
known phenomena, there are plenty of aspects that we do not understand yet and that may suggest
the SM to be most likely a low energy effective theory of spin-1/2 matter fermions interacting
via spin-1 gauge bosons~\cite{Altarelli}.
An excellent candidate to a new theory, able to describe physics at arbitrarily high energies,
is Supersymmetry (SUSY). SUSY is a large class of theoretical models based on the common assumption
that there exist in nature a fermion-boson symmetry.
A comprehensive  SUSY search is almost impossible because of
the large amount of truly independent parameters.
The strategy is then to search for signals suggested by particular models
in which theoretical assumptions are also adopted
to reduce the number of free parameters to a few.
In Supersymmetry fermions can couple to a sfermion and a fermion,
violating lepton and/or baryon number.
To avoid this problem, a discrete multiplicative quantum number,
the $\mathcal{R}$-parity was introduced~\cite{DREINER}:
${\mathcal{R}}\equiv (-1)^{3B+L+2S}$.
SUSY models can be constructed assuming either conservation
or violation of this quantum number (RPV).
%%%%%%%%%%%%%%%%%%%%%%%%%%%%%%%%%%%%%%%%%%%%%%%%%%%%%%%%%%%%%%%%%%%%%%%%%%%%%%%%%%%%%

%%%%%%%%%%%%%%%%%%%%%%%%%%%%%%%%%%%%%%%%%%%%%%%%%%%%%%%%%%%%%%%%%%%%%%%%%%%%%%%%%%%%%
\section{SUSY Searches at Tevatron Collider}

\subsection{Search for third generation scalar quarks}

Search for scalar top squark is particularly interesting as the
strong Yukawa coupling between top/stop and Higgs fields give
rise to potentially large mixing effects and mass splitting.
Such effects can lead the lightest top-squark mass eigenstate $\tilde{t}_{1}$ to be
lighter than the other squarks: $m_{\tilde{t}_{1}} < m_{\tilde{q}}$~\cite{GENSTOP}.
When a set of SUSY parameters such as $A$, $\mu$ and tan($\beta$)~\cite{SUSYPAR}
is suitably tuned, light bottom squarks may also occur.\\
\noindent
Both the CDF and D$\not$O experiments have searched for direct stop quark pair
production: $p \bar{p} \rightarrow \tilde{t}_{1} \bar{\tilde{t}}_{1}$
with $\tilde{t}_{1}$ decaying into the following channels:
$\tilde{t}_{1} \rightarrow  b \tilde{\chi}^{\pm}_{1}$,
$\tilde{t}_{1} \rightarrow  b \ell^{+} \tilde{\nu} $~\cite{stopBCHI}
and
$\tilde{t}_{1} \rightarrow  c \tilde{\chi}^{0}_{1}$~\cite{stopCLSP}.
CDF has also searched for indirect stop quark production
trough  the top quark decay:
$t \rightarrow \tilde{t}_{1}\tilde{\chi}^{0}$
with $\tilde{t}_{1} \rightarrow b \chi^{\pm}_{1}$~\cite{stopDCY}.
\noindent
Searches for direct scalar bottom production
$p \bar{p} \rightarrow \tilde{b}_{1} \bar{\tilde{b}}_{1}$
with the sbottom decaying into:
$\tilde{b}_{1} \rightarrow  b \tilde{\chi}^{0}_{1}$
have been performed from both Tevatron Experiments~\cite{stopCLSP,d0sbottom}.
An overview on such results can be found in~\cite{stopoverview}.
%
% In the next paragraph we will discuss more in details a recent CDF search for
% for RPV stop quark decay:
% $p \bar{p} \rightarrow \tilde{t}_{1} \bar{\tilde{t}}_{1} \rightarrow \tau b \tau b$
%%%%%%%%%%%%%%%%%%%%%%%%%%%%%%%%%%%%%%%%%%%%%%%%%%%%%%%%%%%%%%%%%%%%%%%%%%%%%%%%%%%%%

%%%%%%%%%%%%%%%%%%%%%%%%%%%%%%%%%%%%%%%%%%%%%%%%%%%%%%%%%%%%%%%%%%%%%%%%%%%%%%%%%%%%%
\subsection{Search for RPV stop decays}

CDF searched for a pair produced scalar top squark decaying
via non-zero $\mathcal{R}$-parity violating coupling $\lambda^{'}_{333}$
to  $\tilde{t}_{1} \rightarrow \tau b$~\cite{stopRPV}.
The experimental signature of this process is two $\tau$ leptons and two $b$ quarks
in the final state.
Events have been selected by requiring a lepton ($e$ or $\mu$)
from $\tau \rightarrow \ell \nu_{\ell} \nu_{\tau}$, a hadronically decaying tau lepton
and two jets. The principal background processes are $Z \rightarrow \tau^{+} \tau^{-}$, $W+$jets,
$t \bar{t}$, Drell-Yan and diboson events.
We observed, combining both the muon
$\tilde{t}_{1} \bar{\tilde{t}}_{1} \rightarrow \tau^{+} \tau^{-} b \bar{b} \rightarrow \mu \tau_{h} b \bar{b}+X$
and the electron channel
$\tilde{t}_{1} \bar{\tilde{t}}_{1} \rightarrow \tau^{+} \tau^{-} b \bar{b} \rightarrow e \tau_{h} b \bar{b}+X$,
that no events passed the selection cuts.
This is consistent with the expected SM background of
$1.92 \pm 0.19$ events in the electron channel and $1.13 \pm 0.14$ in the muon channel.
A 95\% {\it C.L.} lower limit on the stop quark mass have been set:
$m_{\tilde{t}_{1}}>$ $119$ GeV/$c^{2}$, for a dominant $\lambda^{'}_{333}$ coupling.
The more recent and competitive result on the lower limit of the stop mass with
this signature comes from ALEPH/LEP experiment~\cite{LEPRPV}.

\subsection{Search for MSSM neutral Higgs}

\begin{figure}[t!]
\hspace{0.2cm}
%\vspace{-1.2cm}
 \begin{minipage}{8.0in}
  \epsfxsize3.74in
  \vspace{-2.8cm}
  \hspace*{-0.2cm}\epsffile{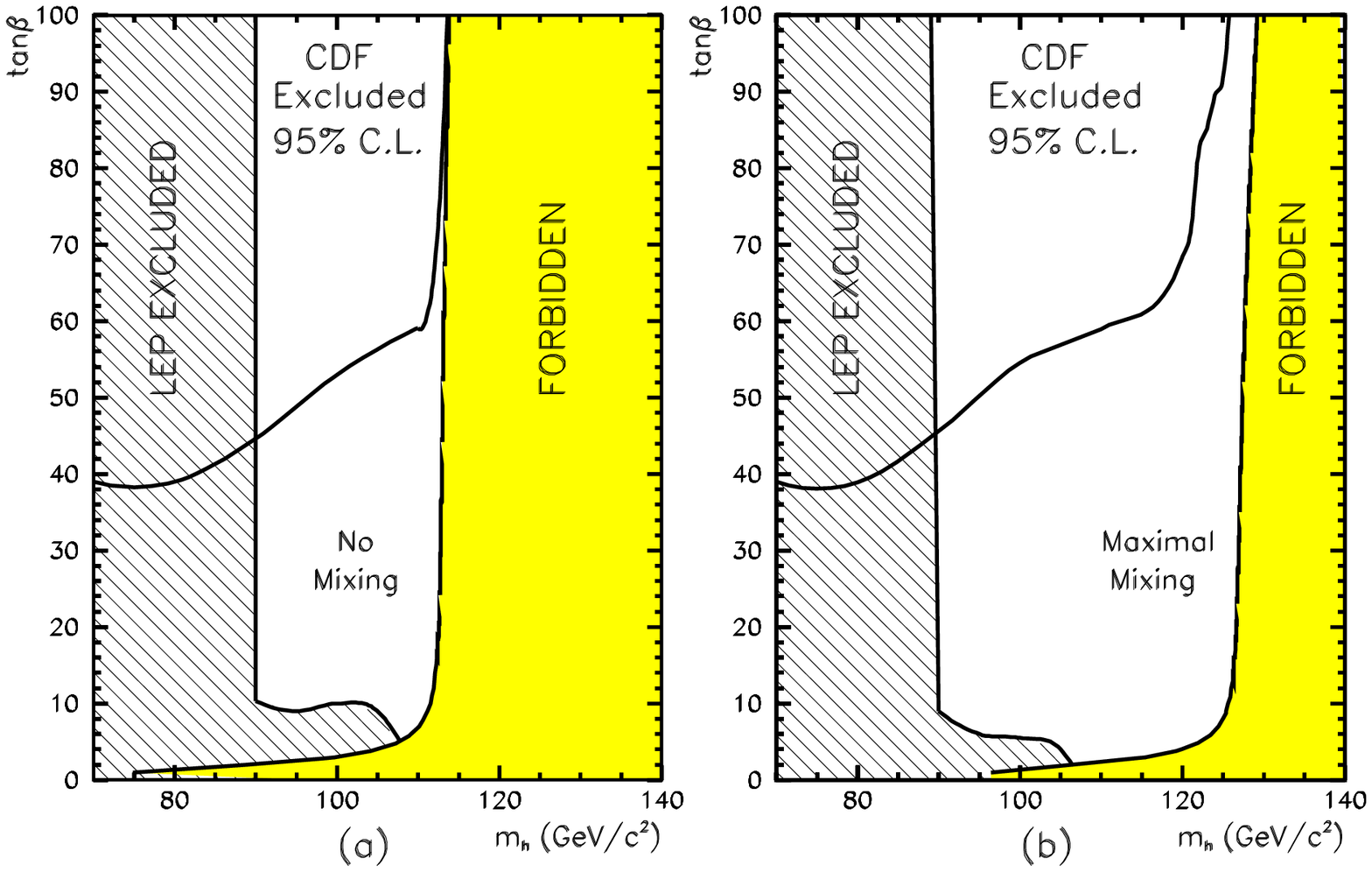}
  \epsfxsize2.5in
  \hspace*{-0.5cm}\epsffile{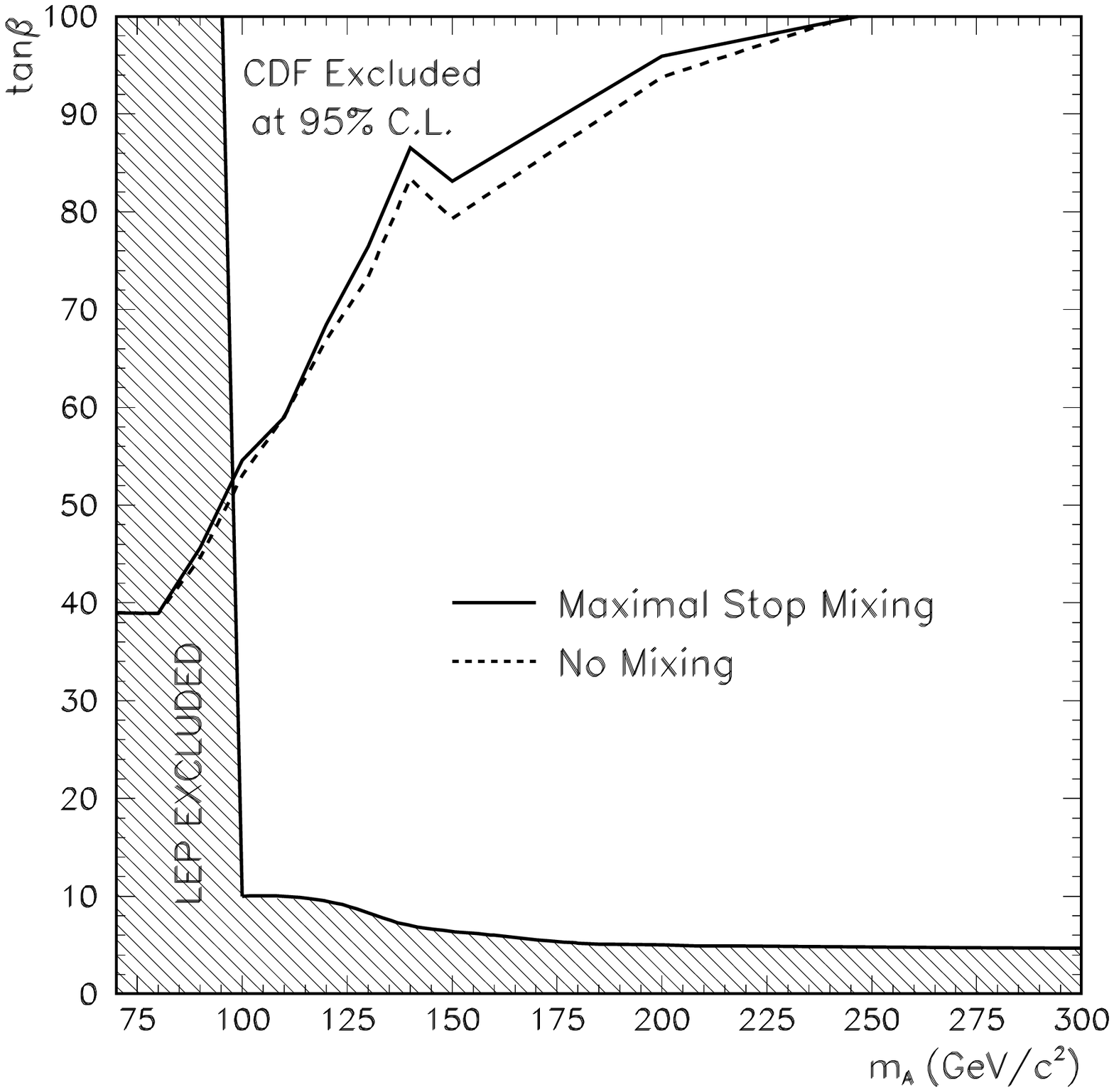}
 \end{minipage}\hfill
\vspace{0.3cm}
 \caption{\it  (left) 95\% C.L. exclusion region in the
$m_{h}$ versus tan($\beta$) plane from CDF MSSM neutral
Higgs search; (right) and $m_{A}$ versus tan($\beta$).}
 \label{Fig:h1h2}
 \end{figure}

The Minimal Supersymmetric Standard Model (MSSM)
predicts five physical Higgs bosons: a charged pair ($H^{+}$, $H^{-}$),
two CP-even scalars ($h^{0}$, $H^{0}$) and a CP-odd ($A^{0}$).
CDF has searched for a neutral MSSM Higgs $\phi$,
%($\phi=$ $h$, $H$, $A$)
where $\phi$ means  $h$ or $H$ or $A$,
produced in association with $b \bar{b}$:
$p \bar{p} \rightarrow b \bar{b} \phi \rightarrow b \bar{b} b \bar{b}$.
The analysis is based on on $91$ pb$^{-1}$
of data corresponding to the Run 1B multijet sample.
With basic parameter choices for both the SUSY scale and the stop mixing,
we obtained a 95\% C.L. on the lower mass value for $\phi$
in a region of SUSY parameter space where: tan$\beta$ $>$ 30.
These results are summarized in  Fig.~\ref{Fig:h1h2}.
%
%for the neutral Higgs sector of the MSSM
%for tan$\beta$ $>$ 30 as shown in Fig.~\ref{Fig:h1h2}.

\subsection{Search for charged Higgs in the top quark decay}

The charged Higgs particle ($H^{\pm}$) may be observed through the
following top quark decay: \,\,$t \rightarrow H^{+} b \rightarrow \tau^{+} \nu b$.
This \,\,\,process \,\,is \,\,favored \,over \,the \,SM one:
\,\,$t \rightarrow Wb$ \,if
m$_{H^{\pm}}$ $<$ $(m_{t} - m_{H})$ in two separate
tan($\beta$) regions:
tan($\beta$)$<$ $1$ and tan($\beta$)$>$  $70$~\cite{chargedHiggs}.
%%%%%%%%%%%%%%%%%%%%%%%%%%%%%%%%%%%%%%%%%%%%%%%%%%%%%%%%%%%%%%%%
\newpage
%%%%%%%%%%%%%%%%%%%%%%%%%%%%%%%%%%%%%%%%%%%%%%%%%%%%%%%%%%%%%%%%
\noindent
Both CDF and  D$\not$O
searched for charged Higgs. In particular
the CDF direct search was performed requiring a high-$P_{T}$ central lepton
($| \eta |<$ $1$, $p_{T}^{\ell}>$ $20$ GeV, $\ell=$ $e$ or $\mu$)
as well as a central $\tau$ lepton with  $p_{T}^{\tau}>$ $15$ GeV,
2 jets and missing transverse energy ($\not\!\!\!E_{\rm T}$)
with significance: $S\,_{\not\!\!\!E_{\rm T}} \equiv \not\!\!\!E_{\rm T} / \sqrt{\sum E_{T}} > 3$ GeV$^{1/2}$.
Better results have been obtained both from
CDF and  D$\not$O performing an indirect search based on the
suppression of SM $t \bar t \rightarrow W^{+} W^{-} b \bar{b}$
decays caused by the presence of the competitive channel $ t \rightarrow H^{+}b$.
Fig.~\ref{Fig:ff2} (left) show the 95\% C.L.
excluded region as a function of tan($\beta$).
%%%%%%%%%%%%%%%%%%%%%%%%%%%%%%%%%%%%%%%%%%%%%%%%%%%%%%%%%%%%%%%%%%%%%%%%%%%%%%%%%%%%%%%%%%%%

%%%%%%%%%%%%%%%%%%%%%%%%%%%%%%%%%%%%%%%%%%%%%%%%%%%%%%%%%%%%%%%%%%%%%%%%
\subsection {Search for gluino pair production using LS top events}

\begin{figure}[t!]
 \hspace{-0.4cm}
%\vspace{-0.4cm}
 \begin{minipage}{8.0in}
  \epsfxsize3.74in
  \hspace*{-0.5cm}\epsffile{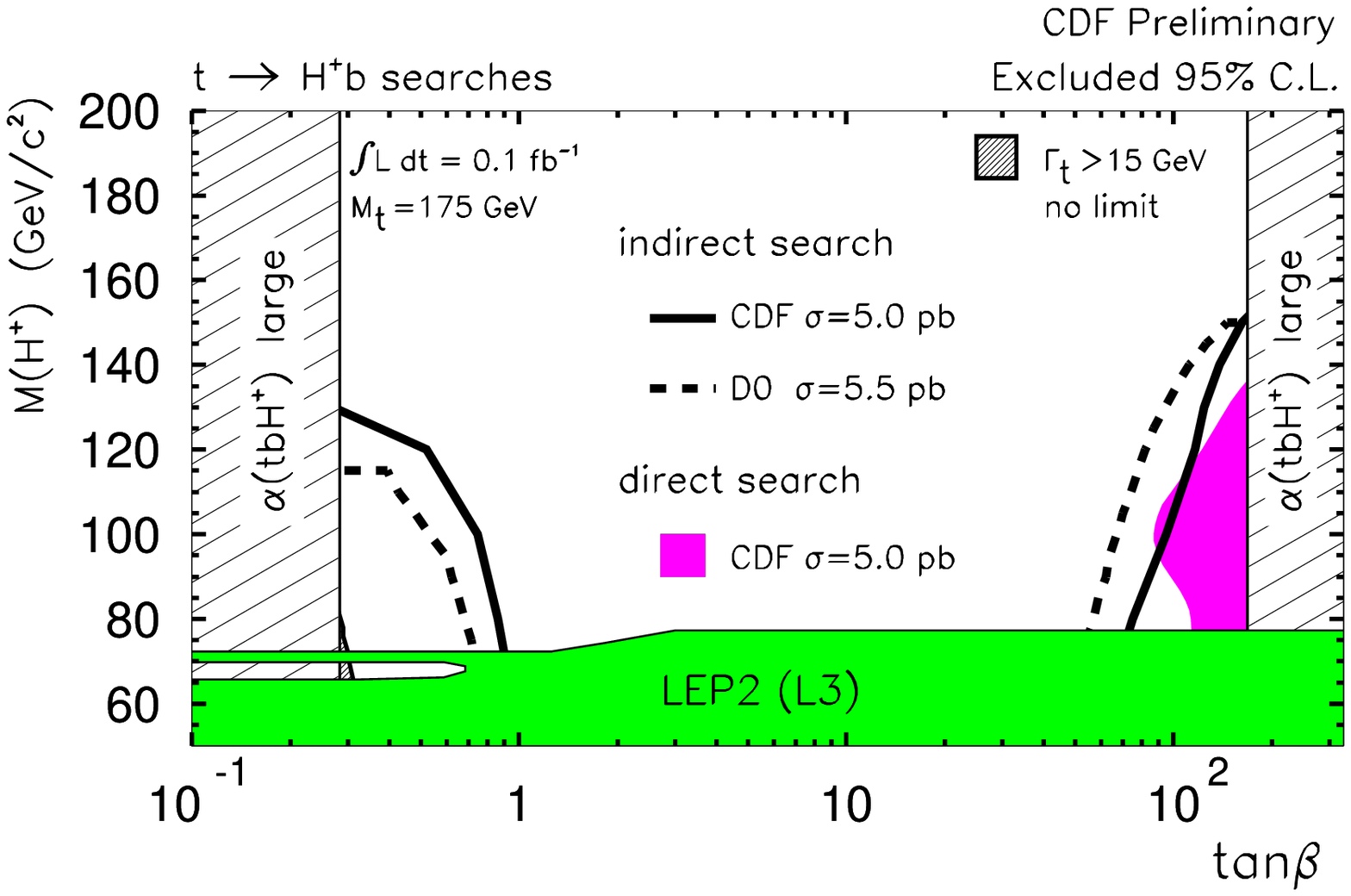}
  \epsfxsize2.8in
  \hspace*{.5cm}\epsffile{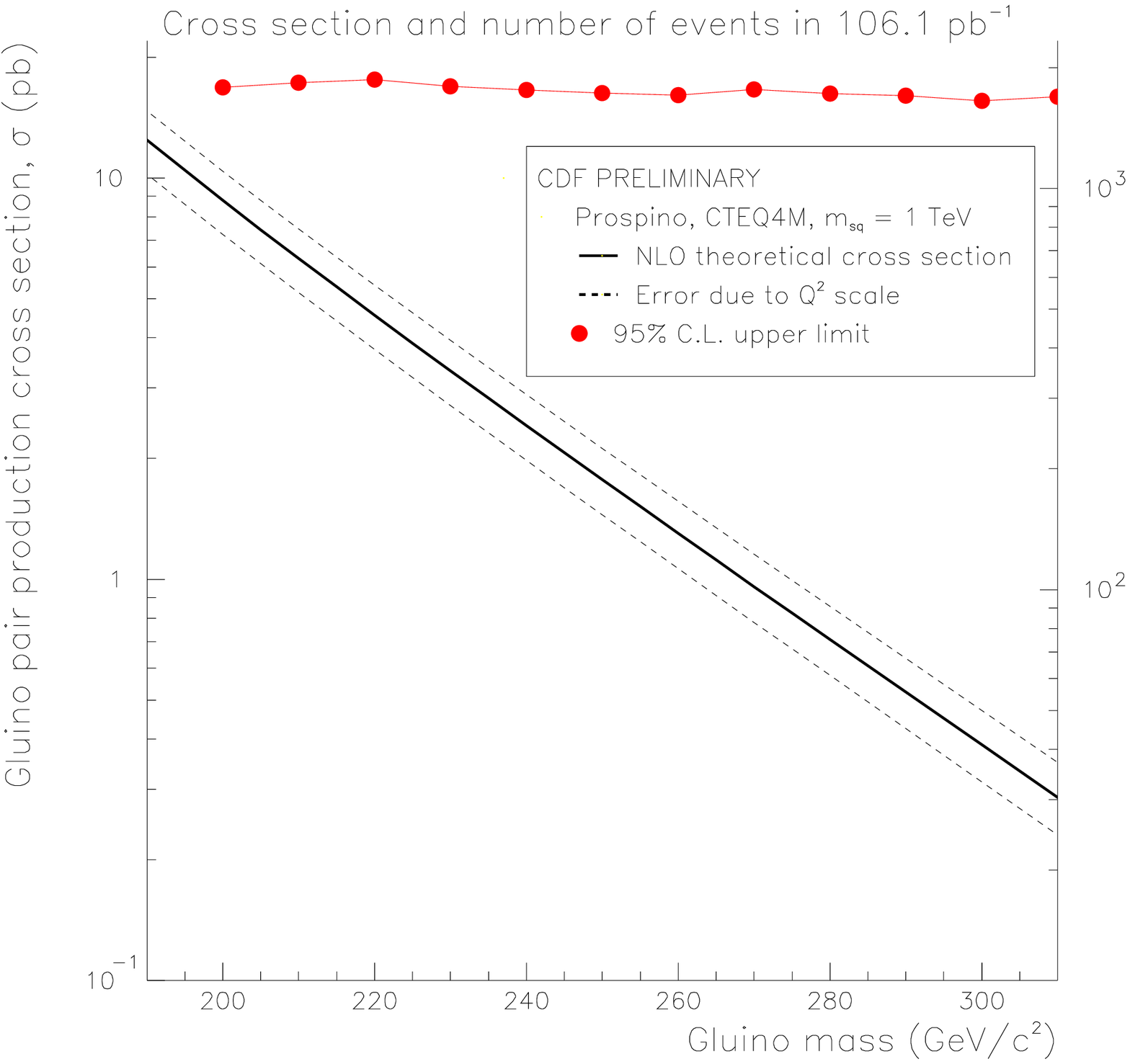}
 \end{minipage}\hfill
 \caption{ \it (left) 95\% C.L. exclusion region in the tan($\beta$) versus
$m_{H^{\pm}}$ plane for the charged Higgs searches from CDF and D$\not$O;
(right) 95\% C.L.upper limit on the gluino production cross section as a
function of $\tilde{g}$ mass using the like-sign top dilepton sample.}
\label{Fig:ff2}
\end{figure}

CDF recently searched for gluino pair production
using like-sign (LS) top events.
The analysis have been performed using $106.1$ pb$^{-1}$ of Run I data.
In the SUSY model under study the scalar top squark
is not only the lightest squark
but also the only one lighter than gluino
and satisfy the condition:
$m_{t} + m_{\tilde{t}_{1}} <$ $m_{\tilde{g}}$.
Therefore $\tilde{g} \rightarrow t \tilde{t}$
is the preferred decay channel and because of the Majorana nature
of gluinos they give rise to LS top quarks from $\tilde{g} \tilde{g}$ decays.
In order to search for such events CDF used the top dilepton events.
The results of this search are shown in Fig.~\ref{Fig:ff2} (right);
no mass limits have been set
due to the presence of three signal events and to the inability to
probe gluino masses in the region close to the
top mass, where the stop mass is forced to be unreasonably light.

\section{Conclusions}
Tevatron Experiments performed extensive searches for physics beyond the Standard
Model using the data collected during the 1992-1996 Run I.
Recent results on such searches have been reported.
No evidence for physics beyond the Standard Model have been found so that
95\% C.L. limit have been set for the different
scenarios described in the present paper.
\,With \,the \,\,Run II \,upgrades, \,\,providing \,a \,higher \,\,acceptance \,\,and

%%%%%%%%%%%%%%%%%%%%%%%%%%%%%%%%%%%%%%%%%%%%%%%%%%%%%%%%%%%%%%%%
\newpage
%%%%%%%%%%%%%%%%%%%%%%%%%%%%%%%%%%%%%%%%%%%%%%%%%%%%%%%%%%%%%%%%
\noindent
higher luminosity, it will
be possible to make important progress in the search for new phenomena.

\section{Acknowledgments}
We would like to thank the organizers of the
{\it XIII Convegno sulla Fisica al LEP} for the excellent conference and
for their kind hospitality.
%

%%%%%%%%%%%%%%%%%%%%%%%%%%%%%%%%%%%%%%%%%%%%%%%%%%%%%%%%%%%%%%%%%%%%%%%%
% End of the paper
\end{document}